# Excitation of magnon accumulation by laser clocking as a source of long-range spin waves in transparent magnetic films

M. Jäckl[1], V.I. Belotelov[2,3], I.A. Akimov[1,4], I.V. Savochkin[2], D.R. Yakovlev[1,4], A.K. Zvezdin[3,5], M. Bayer[1,4]

[1]*Experimentelle Physik 2, TU Dortmund, D-44221 Dortmund, Germany*
[2]*Lomonosov Moscow State University, 119991 Moscow, Russia*
[3]*Russian Quantum Center, Skolkovo, 143025 Moscow, Russia*
[4]*Ioffe Institute, Russian Academy of Sciences, 194021 St. Petersburg, Russia*
[5]*Moscow Institute of Physics and Technology, Moscow Region 141700, Russia*

**ABSTRACT**

Optical tools are of great promise for generation of spin waves due to the possibility to manipulate on ultrashort time scales and to provide local excitation. However, a single laser pulse can inject spin waves only with a broad frequency spectrum, resulting in a short propagation distance and low amplitude. Here we excite a magnetic garnet film by a train of fs-laser pulses with 1 GHz repetition rate so that pulse separation is smaller than decay time of the magnetic modes which allows to achieve collective photonic impact on magnetization. It establishes a quasi-stationary source of SWs, namely a coherent magnon accumulation ("magnon cloud"). This approach has several appealing features: (i) the source is tunable; (ii) the SW amplitude can be significantly enhanced; (iii) the spectrum of the generated SWs is quite narrow that provides longer propagation distance; (iv) the periodic pumping results in almost constant in time SW amplitude up to 100 μm away from the source; and (v) the SW emission shows a pronounced directionality. These results expand the capabilities of ultrafast coherent optical control of magnetization and pave a way for applications in data processing, including the quantum regime. The quasi-stationary magnon accumulation might be also of interest for the problem of magnon Bose-Einstein condensate.

**INTRODUCTION**

Recent research on spin waves (SW) is increasingly driven by their unique linear and non-linear properties as well as anticipated applications in telecommunications, image processing and even quantum computations [1-5]. SWs are launched if in a magnetically ordered material the magnetization is pushed out of equilibrium. Usually, this is achieved by microwaves generated by an antenna in close vicinity of the sample [6]. However, particular applications require a strong locality of the excitation and a specific distribution of spins in time and space, created on time scales much shorter than any decay time. For example, quantum information processing necessitates addressing a qubit by a magnetic field with a submicron gradient [5]. This challenge might be solved if the magnetic system is disturbed by ultra-short laser pulses, that can be focused microscopically [7-25]. Then the instantaneous impact of the laser pulse on the magnetization occurs only within the illumination spot with potentially sub-wavelength resolution if plasmonic nanogeometries are used [26,27].

Among the various mechanisms for optical pumping of magnetization in ferromagnets, the inverse Faraday effect is of particular importance [8,21-24]. Here circularly polarized light affects the medium magnetization, as if an effective magnetic field $\mathbf{H}_F \sim [\mathbf{E} \times \mathbf{E}^*]$ would act, where $\mathbf{E}$ is the electric field of the light wave [28]. As a result, $\mathbf{H}_F$ is directed along the light wave vector. In ferromagnets it originates from stimulated Raman scattering on magnons. The inverse Faraday effect was observed in pump-probe experiments [21-24], where the magnetization dynamics is triggered by a pump and subsequently monitored by a delayed probe pulse of low intensity. The authors of Refs. [8-10] managed to demonstrate magnetostatic SWs in iron garnets in that way.

Almost all pump-probe studies so far were conducted in the single pump pulse regime where the magnetization oscillations decay before the subsequent pulse arrives. Double-pump coherent magnetization control was demonstrated in Ref. [21]. Recently, the magnetization precession was excited by a sequence of picosecond acoustic pulses [29]. However, so far there were no studies of the impact of a virtually infinite sequence of pump pulses exciting the sample at high rate such that upon pulse arrival the magnetization still shows the coherent dynamics induced by the preceding pulse.

Here we excite transparent magnetic films with a train of fs-laser pulses hitting periodically the sample with a period of 1 ns. This pulse separation is comparable to or even shorter than the decay time of the magnetization modes. Being periodic in time, the excitation generates a $\mathbf{H}_F(t)$ in the illuminated area which represents a quasi-continuous source of SWs showing several appealing features, namely directionality of the SWs as well as enhanced amplitudes for specific SW frequencies, providing frequency selectivity.

**MATERIALS AND METHODS**

The experiments here are conducted on monocrystalline magnetic films of bismuth-substituted iron-garnet: a 5-µm-thick film with chemical composition $(Bi_{0.9}Lu_{1.4}Tm_{0.4}Y_{0.2}Sm_{0.1})(Fe_{4.7}Ga_{0.3})O_{12}$ (sample-1, Figs. 1,4) and 4-µm-thick film of composition $(Bi_{0.8}Lu_{2.2})Fe_5O_{12}$ (sample-2, Figs. 2,3). Both films were grown by liquid phase epitaxy on gadolinium gallium garnet (GGG) with crystallographic orientation (111).

The magnetization precession is excited and detected using a pump-probe technique based on asynchronous optical sampling (ASOPS). Two independent laser oscillators for the circularly polarized pump and the linearly polarized probe beams emit 50 fs pulses with a center wavelength of $\lambda \approx 800$ nm at a rate of about 1 GHz (Fig. 1a). The repetition frequencies of the oscillators are synchronized to each other with a small offset of 2 kHz. As a result, the relative time delay between pump and probe pulses is repetitively ramped from zero to 1 ns within a scan time of 5 µs. The pump and probe beams are focused onto the sample using a single reflective microscope objective with a magnification factor of 15 comprising 4 sectors through which the light can enter and exit. Pump and probe beams are incident at an angle $\beta = 17°$ in planes orthogonal to each other (the *XZ*- and *YZ*-planes in Fig. 1a). The pump pulses with energy of around 50 pJ are focused into a spot size of about 10 μm diameter while the probe pulses with energy of about 3 pJ are focused into a spot of about 7 μm diameter.

Variations of the out-of-plane magnetization component, i.e. the component along the *z*-axis, are detected by the Faraday rotation angle $\Phi$ of the probe beam with variable time delay relative to the pump. The angle $\Phi$ is measured using a polarization bridge which comprises a Wollaston prism and a 10-MHz balanced photodetector. The amplified differential signal is sent to a high speed multi-channel digitizer triggered by the ASOPS system at the frequency of 2 kHz. The resulting Faraday rotation angle is given by $\vartheta(t) = 4dU(t)/U_{dc}$ where $dU(t)$ is the time-resolved differential signal and $U_{dc}$ is the average intensity measured with one of the photodiodes. The magnetization precession angle $\theta$ can be deduced from $\Phi$ (Eq. (S3), Supplementary). The external magnetic field up to 2500 Oe is applied in-plane, along the *x*-axis, of the sample using an electromagnet. All measurements are performed at room temperature.

For numerical calculation of spin waves amplitude and dispersion (see Figs. 4b,4c) we used the following magnetic parameters of the 5-μm-thick film: gyromagnetic ratio $\gamma = 1.76 \cdot 10^7$ rad·Oe$^{-1}$·s$^{-1}$, magnetization of saturation $4\pi M_s = 1000$ Oe, uniaxial anisotropy constant $K_u = 3.0 \cdot 10^4$ erg·cm$^{-3}$ and cubic anisotropy constant $K_1 = -1.2 \cdot 10^3$ erg·cm$^{-3}$.

**RESULTS AND DISCUSSION**

**Collective photonic impact on magnetization**

The train of circularly polarized pump pulses excites a periodic magnetization dynamics (Fig. 1b) due to the inverse Faraday effect as confirmed by the fact that switching the pump helicity from $\sigma^+$ to $\sigma^-$ changes the precession phase by $\pi$. Sweeping the external magnetic field modifies the amplitude $\theta_0$, phase $\zeta$ and frequency $\omega$ of the oscillations. In the time interval of $T = 1$ ns between two consecutive laser pulses the oscillations can be approximately described by a decaying sine function: $\theta = \theta_0 e^{-t/\tau} \sin(\omega t + \zeta)$ (Fig. 1c), where $\tau$ is the decay time. Small deviations of the measured data from the fit curves are caused by excitation of SWs. Strikingly, the precession amplitude shows resonances for particular magnetic fields, at which the oscillation frequency is a multiple of the laser repetition rate, corresponding to the relation: $\omega T/2\pi = N$, where $N$ is an integer (Fig. 1d). Accordingly for $N = 4, 5, 6$, and 7 the resonances appear at $H = 1400, 1726, 2080$, and 2427 Oe, respectively. At such a resonance, $\theta_0$ exceeds its minimal value by 60% and $\zeta$ vanishes (see blue curve at $H = 1400$ Oe in Fig. 1c) so that there is no phase shift which is a result from spin synchronization: the oscillations at each excitation cycle are synchronized with each other.

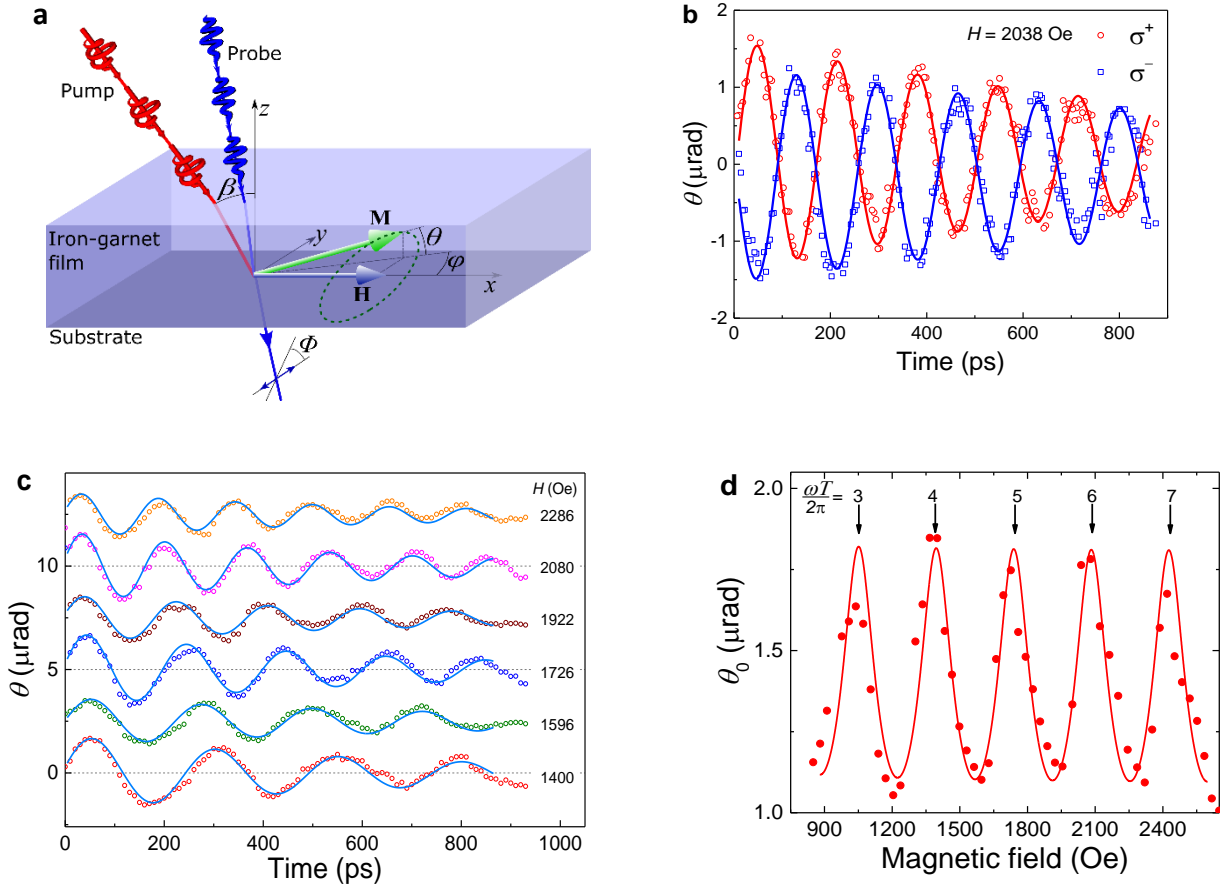

**Figure 1. Excitation of magnetization precession in an iron-garnet film by a train of fs optical pulses. a,** Experimental scheme. The periodic laser pulses are incident on the magnetic film. The probe beam is detected from the substrate side of the samle. **b**, Comparison of magnetization oscillations excited by a $\sigma^+$ and $\sigma^-$ circularly polarized pump (red circles and blue squares, respectively). Solid curves are fits with a decaying sine function. The fit parameters are given in Table SI (Supplementary). $H = 2038$ Oe. **c**, Magnetization oscillations at different magnetic fields (thick curves) corresponding to maximal and minimal amplitudes. Fits to the data by decaying sines are given by the thin blue curves. The incident light is $\sigma^+$-polarized. Fit parameters are given in Table SII (Supplementary). **d**, Measured precession amplitude versus magnetic field strength (circles). The solid curve is calculated after Eq.(3) using the parameters $h = 6.2$ Oe, $\alpha = 0.049$ ($\tau = 0.77$ ns at $H = 1500$ Oe) to fit the experimental data. Experimental results are presented for sample-1.

For some samples the magnetization precession does not occur on a single frequency, see Fig. 2a. For all magnetic fields shown there the signal contains a fast decaying component, termed mode-A in the following, with a decay time of about 0.7 ns, while for some resonant magnetic fields (e.g. for $H = 1719$ Oe or 2064 Oe) an additional, slowly decaying component, mode-B, can be clearly identified at probe-pump delays exceeding 0.7 ns. Mode-B disappears completely for even slight detuning from the resonant fields (compare the oscillations at $H = 1719$ Oe and 1735 Oe). By subtracting the off-resonant signals from the resonant ones, one can isolate mode-B which almost does not decay in the 1 ns time interval. Four consecutive mode-B resonances corresponding to $N = 8, 9, 10$ and 11 are shown in Fig. 2b.

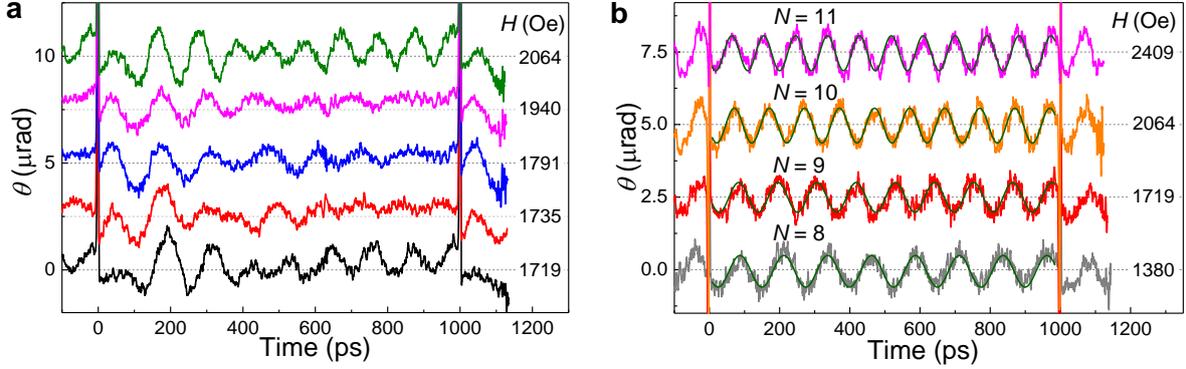

**Figure 2. Excitation of magnetization oscillation showing two precession modes in an iron-garnet film by a laser pulse train. a**, Magnetization oscillations at different magnetic fields. **b**, Consecutive resonances of the slowly decaying precession mode-B for different magnetic fields. The green thin lines are fits with decaying sine functions. The fit parameters are given in Table SIII (Supplementary). The experimental data are presented for sample-2.

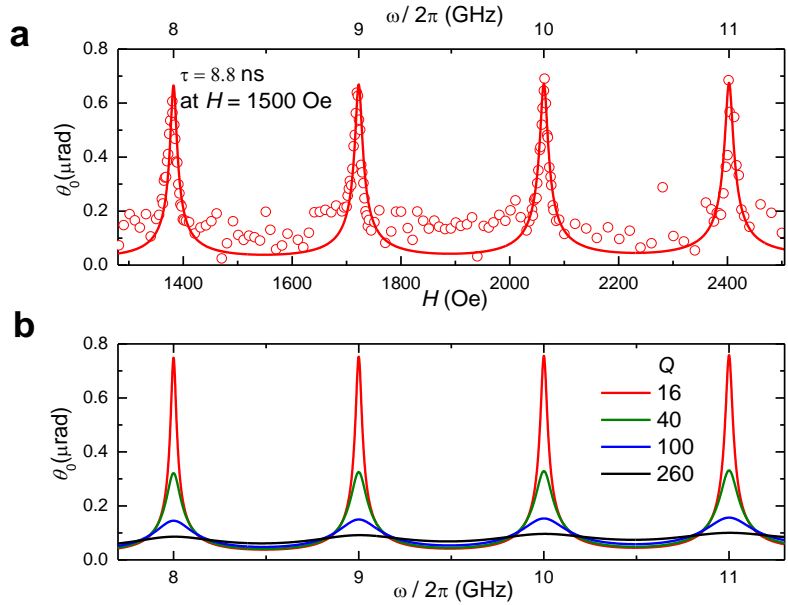

**Figure 3. Excitation of magnetization oscillation mode of high quality factor (mode-B). a**, Measured precession amplitude in sample-2 versus magnetic field (the open circles). Solid curve is calculated by solving Eq. (S11) (Supplementary) with $h = 0.8$ Oe, and $\alpha = 0.0017$ ($\tau = 8.8$ ns at $H = 1500$ Oe) to fit the experimental data. **b**, Calculated precession amplitude versus $\omega_0$ for modes of different quality factors (and therefore different values of $\tau$: 0.6 ns (black), 1.5 ns (blue), 4 ns (green), and 10 ns (red) at $H = 1500$ Oe). $h = 0.8$ Oe.

The amplitude of mode-A behaves quite similar to the mode discussed above in Figs. 1 and 2. However, the mode-B amplitude changes with magnetic field much more prominently (Fig. 3a). In resonance where its frequency is a multiple of the excitation rate the mode-B amplitude is 5 times above the noise level and comparable to the mode-A amplitude. Remarkably, in the single pulse excitation regime, i.e. when the sample is excited at much lower repetition rate ($1/T$ = 80 MHz), the amplitude of mode-B is much smaller than that of mode-A and, therefore, cannot be resolved.

**Analysis of the observed phenomena**

A quantitative understanding of the observations can be obtained from the dynamical equation for the magnetization. The excitation by optical pulses generally leads to a spatial and temporal pattern of the magnetization $\mathbf{M}(\mathbf{r}, t)$. The magnetization dynamics of these modes is described by the Landau-Lifshitz-Gilbert equation [31]:

$$\frac{d\mathbf{M}_q}{dt} = -\gamma \mathbf{M} \times \mathbf{H}_{eff} + \frac{\alpha}{M} \mathbf{M} \times \frac{d\mathbf{M}}{dt}, \quad (1)$$

where $\gamma$ is the gyromagnetic ratio and $\alpha$ is the Gilbert damping constant. The effective magnetic field $\mathbf{H}_{eff}$ in this equation is determined by the variational derivative of the sample free energy $W$ with respect to the magnetization: $\mathbf{H}_{eff} = -\delta W/\delta \mathbf{M}$. The free energy is the sum of several contributions: $W = W_Z + W_d + W_a + W_F$, where $W_Z$ is the Zeeman energy in the external magnetic field $\mathbf{H}$, $W_d$ and $W_a$ are the demagnetization and magnetic anisotropy energies, respectively. The last term, $W_F$, takes into account the inverse Faraday effect: $W_F = -\mathbf{M} \cdot \mathbf{H}_F$. When a laser pulse propagates through the magnetic film at an angle $\beta$ relative to its normal and $\mathbf{H}$ has components both in the sample and the light incidence planes, Eq. (1) leads to (Supplementary):

$$\frac{\partial^2 \theta}{\partial t^2} + \alpha(\omega_0 + \omega_e)\frac{\partial \theta}{\partial t} + (\omega_0 + \omega_{F\perp}(t))(\omega_e + \omega_{F\perp}(t))\theta = \omega_0 \omega_{F\|}(t), \quad (2)$$

where $\omega_0 = \gamma H$, $\omega_e = \gamma(H + 4\pi M + H_a)$, $\omega_{F\|}(t) = \gamma H_F(t)\cos\beta$, and $\omega_{F\perp}(t) = \gamma H_F(t)\sin\beta$.

If magnetic losses are neglected and $\beta \to \pi/2$, Eq. (2) becomes isomorphic to the Schrödinger equation with a periodic potential $V(t) = -\gamma H_F(t)(2\omega_0 + \omega_a)$, and the problem is reduced to the Kronig-Penney model with eigenfrequencies $\omega_0(\omega_0 + \omega_a)$. For relatively large $H_F$ band gaps in the oscillation spectrum appear so that appealing effects such as the parametric generation of magnetization oscillations become possible.

In our experimental conditions the laser pulses propagate through the film almost along its normal so that $\beta \ll 1$. In this case the right hand part of Eq. (2) becomes prominent and the inverse Faraday effect acts on the magnetization as an external periodic force. Since the optical pulse duration $\Delta t \ll T, \omega_H^{-1}$, the following representation of this force can be used: $\omega_{F\|}(t) = \gamma \Delta t \sum_{m=0}^{+\infty} h_m \delta(t - mT)$, where $h_m$ is the amplitude of the $m$-th pulse in $H_F$.

The problem can be solved using the Green function formalism: $\theta(t) = \frac{1}{\omega}\sum_{m=0}^{+\infty} h_m G(t + mT)$, where the Green function $G(t) = \omega^{-1}\sin\omega t\, e^{-t/\tau}$ for $t \geq 0$, $G(t) = 0$ for $t < 0$, and $\tau =$ $2/(\alpha(\omega_0 + \omega_e))$. If the train of fs-pulses is uniform, i.e. $h_m = h = $ const, then the oscillation regime becomes quasi-stationary and $\theta(t)$ is a periodic function of $T$: $\theta(t) = \theta_0 \sin(\omega t + \zeta) e^{-t/\tau}$ within a single interval between two pump pulses, i.e. for $mT < t < (m+1)T, m \gg 1$. Here $\omega = \omega_0\sqrt{1 - (\omega_0\tau)^{-1}} \approx \omega_0$. For relatively large optical losses so that $e^{-T/\tau} \ll 1$, the amplitude and phase of the magnetization oscillation are given by (Supplementary):

$$\begin{cases} \theta_0 = \frac{\omega_0}{\omega} \gamma h \Delta t \left(1 + e^{-T/\tau} \cos \omega T\right) \\ \zeta = e^{-T/\tau} \sin \omega T \end{cases}. \qquad (3)$$

Accordingly, the precession amplitude shows a resonance when the oscillations are synchronized with the laser pulses, i.e. at $\omega T = 2\pi N$. At these resonances $\zeta = 0$, i.e. the oscillations start with $\theta = 0$ at the moment of subsequent pulse arrival, which is in good agreement with the experiment (Fig. 1c). The ratio of the maximal and minimal amplitudes is given by

$$\theta_{max}/\theta_{min} = \tanh(2\tau/T) = \tanh(4Q/\omega_0 T). \qquad (4)$$

Therefore, the amplitude is increased most strongly for a magnetization mode of high quality factor $Q$ (Fig. 4b). The enhancement factor relative to the amplitude $\theta_s$ of the oscillations excited by a single pump pulse is $\theta_{max}/\theta_s = \left(1 - e^{-T/\tau}\right)^{-1}$, showing the resonant enhancement if $T \ll \tau$. Fitting the observed resonances at $H = 1500$ Oe with the theoretical forms as shown in Figs. 1d and 4a gives $\tau = 0.77$ ns and $\tau = 8.8$ ns for mode-A and B, respectively. These values correspond to enhancement factors of 1.4 and 9.3 for these two modes.

**Observation of tunable spin waves**

Let us now shift the probe beam with respect to the pump to trace the SWs. The oscillating signals observed for different pump-probe separations confirm propagation of the SWs in the direction along field **H** (Fig. 4a). SWs are detectable for spatial shifts up to 100 μm (open red circles in Fig. 4b). Notably, this distance is twice larger than for SWs excited in a single pulse regime. On the contrary, in the orthogonal direction the SWs are only observable within 10 μm from the pump beam center (filled orange circles in Fig. 4b). One more remarkable point is that the SW decay rate diminishes with increasing pump-probe distance. For propagation distances larger than 10 μm the rate undercuts 0.3 ns[-1] (Fig. 4b) so that the oscillations almost do not decay in time in between the pump pulses (Fig. 4a).

All these unique features of the periodic pumping of SWs are related again to the synchronization of oscillations and also involve the specific character of the SW dispersion. Since the Gilbert damping in iron-garnets is relatively small: $\alpha \sim 10^{-3}$ which corresponds to $\tau \sim 100$ ns, the decay time is mainly determined by the spread of SW energy away from the excitation spot and the dephasing of SW modes of different frequency due to the SW dispersion. While the former is inevitable, the latter may be tailored as in our case by the periodic pumping, because the repeated application of the pump pulses narrows the spectrum of generated SWs. Indeed, a single pump pulse with spot radius $r$ excites SWs with wave numbers $k < 1/r$ corresponding to frequencies in the range $\omega(1/r) < \omega < \omega(0)$ (for the backward magnetostatic SWs excited in our experiments) (Fig. 4c). However, the train of pulses singles out frequencies that are multiples of the laser repetition rate ($1/T$) from this range (see red dashed line for $\omega = 6\pi/T$). Since the SW spectrum becomes narrower thereby, the dephasing decreases and the SWs propagate over longer distances.

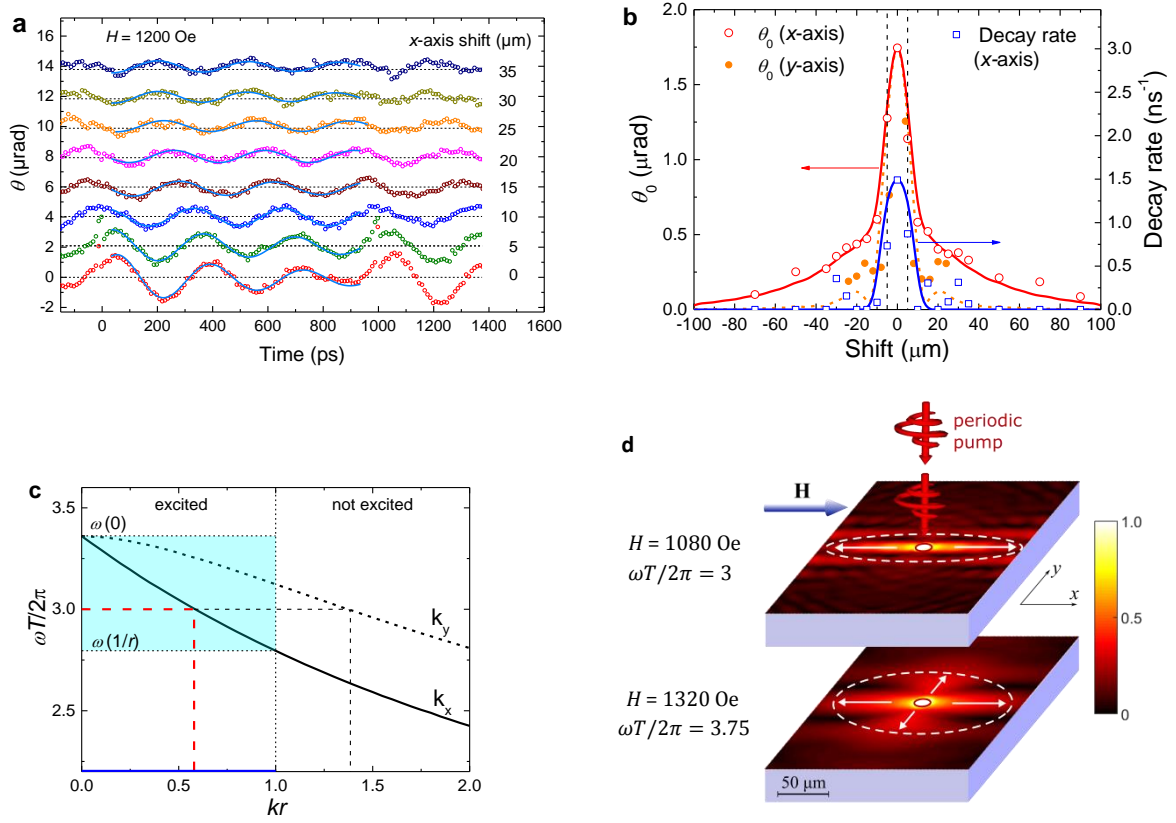

**Figure 4. Generation of spin waves by a cloud of periodically pumped, coherently oscillating spins. a,** Magnetization oscillations observed in the sample-1 in the probe beam shifted by $5-35$ μm relative to the pump beam along the external magnetic field. Solid curves are fits with a decaying sine function. The fit parameters are given in Table SII (Supplementary). $H = 1060$ Oe. **b,** Oscillation amplitude (red, orange) and decay rate (blue) at different distances between pump and probe for the probe shifted along the *x*-axis (open symbols and solid lines) and the *y*-axis (filled symbols and dashed lines). The experimental data are shown by the symbols, while the calculation results are given by lines. **c,** Calculated dispersion of the SWs propagating along the *x*- and *y*-directions. The blue region indicates the range of *k* generated by a single laser pulse with spot radius *r*. The dispersion was calculated using the theoretical model in Ref. [30]. **d,** Calculated distribution of $m_z$ in SWs launched at $H = 1060$ Oe ($\omega T/2\pi = 3$), $H = 1320$ Oe ($\omega T/2\pi = 3.75$). In the former case SWs are generated in a narrow region along **H** (phenomenon of SW directionality), while in the latter one SWs are generated in perpendicular direction as well.

As the SW spectrum is anisotropic with respect to **H** so that $\omega(k_x, 0) \neq \omega(0, k_y)$ (see Fig. 4c) the laser pulse train excites SWs along different directions with various efficiencies. For example, at $H = 1200$ Oe a multiple of the pulse repetition rate is within the frequency window of the excitable SWs along the *x*-axis (for $= 0.6 < 1$ $\omega T/2\pi = 3$, Fig. 2c) while there is no multiple in the SW frequency range along the *y*-axis (dashed curve in Fig. 2c). As a result, SWs can be traced along the *x*-axis over distances of about 100 μm, but they can hardly be observed along the *y*-axis even for small distances from the pumped spot. Therefore, the periodic pumping introduces an additional potentiality as it excites SWs along specific directions which may be altered by changing the pulse repetition rate and the external magnetic field (Fig. 4d).

Finally, the phenomenon of the vanishing time decay of SWs at sufficiently large distances from the pump beam can be qualitatively explained by the local balance of the energy arriving from the optically excited area and the energy carried away by the SWs. The latter depends on the SW dispersion. The experimental data on the oscillation amplitude and the decay rate as function of separation between pump and probe are in an excellent agreement with calculations based on the SW dispersion (solid curves in Fig. 2b, Supplementary).

**CONCLUSIONS**

Summarizing the experimental data and their analysis, we have demonstrated that a train of optical pump pulses can drive the magnetization within the illuminated area thus exciting magnons. The magnons become spread by propagation from the excitation spot and their superposition is observable as SWs. Therefore, the excited area can be considered as a coherent magnon accumulation ("cloud") which serves as source of SWs with parameters controllable by the details of the experimental protocol such as the laser illumination or the magnetic field.

The magnon cloud has several outstanding properties. In particular, the amplitude of the oscillations in this cloud can be significantly increased if the magnetization oscillation frequency is tuned by the magnetic field of moderate strength such that it becomes synchronized with the laser pulses. The enhancement factor can be as large as 9.3 for a magnetic mode of high quality factor ($Q \sim 232$). Such enhancement has allowed us to observe a magnetization mode that is generally obscured in experiment using single pulse excitation by other modes of lower quality factor. Our technique therefore can be applied as highly sensitive spectroscopy tool for resolving magnetization precession modes. Remarkably, the decay rate of the precession amplitude strongly depends on the distance from the optical pump. At some distance the rate becomes so small that the oscillation hardly undergoes a decay anymore. Moreover, the periodic pumping additionally brings two important features for SW propagation: it significantly increases their propagation distance and provides also the possibility to tune directionality of propagation.

Generally, optical excitation of magnetic oscillations represents a detailed control tool of the magnetization distribution at sub-terahertz time scales. In combination with focusing by plasmonic antennas, also the spatial distribution can be tailored on subwavelength scales which is of prime importance for quantum information processing based on SWs [4]. The tunable, quasi-stationary magnon cloud as spin wave source might significantly broaden the functionality of this approach. The next step forward will be implementation of plasmonic structures to further enhance the magnetization precession by concentrating the optical fields in a nanometer thick magnetic film with a spot size of less than 100 nm in diameter [26]. Indeed, while a huge plasmonics-mediated increase of the direct magneto-optical effects was demonstrated recently [32], the plasmonic boost of the inverse magneto-optical effects is still waiting for its practical implementation [16].

## ACKNOWLEDGEMENTS

MJ, IAA, DRY and MB acknowledge financial support by the Deutsche Forschungsgemeinschaft (Project No. AK40/7-1 and TRR 160), while VIB and AKZ acknowledge financial support by the Russian Science Foundation (Project No. 14-32-00010) for conceiving the experiments, fabrication of the samples and theoretical part of the work.

## REFERENCES

1. S. O. Demokritov, A. N. Slavin, *Magnonics: From fundamentals to applications* (Springer, Heidelberg, 2013).
2. A. V. Chumak, V. I. Vasyuchka, A. A. Serga, B. Hillebrands, *Magnon spintronics,* Nature Phys. **11**, 453–461 (2015).
3. B. Lenk, H. Ulrichs, F. Garbs, M. Münzenberg, *The building blocks of magnonics*, Phys. Rep. **507**, 107–136 (2011).
4. Y. Tabuchi, S. Ishino, A. Noguchi, T. Ishikawa, R. Yamazaki *et al. Coherent coupling between a ferromagnetic magnon and a superconducting qubit*, Science **349**, 405–408 (2015).
5. J. H. Wesenberg, A. Ardavan, G. A. D. Briggs, J. J. L. Morton, R. J. Schoelkopf *et al. Quantum computing with an electron spin ensemble,* Phys. Rev. Lett. **103**, 070502 (2009).
6. V. E. Demidov, M. P. Kostylev, K. Rott, P. Krzysteczko, G. Reiss *et al. Excitation of microwaveguide modes by a stripe antenna,* Appl. Phys. Lett. **95**, 112509 (2009).
7. X. Zhang, C. Zou, N. Zhu, F. Marquardt, L. Jiang *et al. Magnon dark modes and gradient memory,* Nat. Commun. **6**, 8914 (2015)
8. T. Satoh, Y. Terui, R. Moriya, B. A. Ivanov, K. Ando *et al. Directional control of spin-wave emission by spatially shaped light*, Nature Photon. **6**, 662–666 (2012)
9. I. Yoshimine, T. Satoh, R. Iida, A. Stupakiewicz, A. Maziewski *et al. Phase-controllable spin wave generation in iron garnet by linearly polarized light pulses*, J. Appl. Phys. **116**, 043907 (2014).
10. S. Parchenko, A. Stupakiewicz, I. Yoshimine, T. Satoh, A. Maziewski, *Wide frequencies range of spin excitations in a rare-earth Bi-doped iron garnet with a giant Faraday rotation*, Appl. Phys. Lett. **103**, 172402 (2013).
11. M. van Kampen, C. Jozsa, J. T. Kohlhepp, P. LeClair, L. Lagae *et al. All-optical probe of coherent spin waves*, Phys. Rev. Lett. **88**, 227201 (2002)
12. Y. Au, M. Dvornik, T. Davison, E. Ahmad, P. S. Keatley *et al. Direct excitation of propagating spin waves by focused ultrashort optical pulses,* Phys. Rev. Lett. **110**, 097201 (2013)
13. B. Lenk, G. Eilers, J. Hamrle, M. Münzenberg, *Spin-wave population in nickel after femtosecond laser pulse excitation*, Phys. Rev. B **82**, 134443 (2010)
14. E. Beaurepaire, J. Merle, A. Daunois, J. Bigot, *Ultrafast spin dynamics in ferromagnetic nickel,* Phys. Rev. Lett. **76**, 4250–4253 (1996)
15. A. M. Kalashnikova, A. V. Kimel, R. V. Pisarev, V. N. Gridnev, P. A. Usachev *et al. Impulsive excitation of coherent magnons and phonons by subpicosecond laser pulses in the weak ferromagnet FeBO3*, Phys. Rev. B **78**, 104301 (2008)

16. V. I. Belotelov, A. K. Zvezdin, *Inverse transverse magneto-optical Kerr effect*, Phys. Rev. B **86**, 155133 (2012).
17. J. Bigot, M. Vomir, E. Beaurepaire, *Coherent ultrafast magnetism induced by femtosecond laser pulses*, Nature Phys. **5,** 515–520 (2009).
18. B. Koene, M. Deb, E. Popova, N. Keller, Th. Rasing *et al*, *Excitation of magnetic precession in bismuth iron garnet via a polarization-independent impulsive photomagnetic effect*, Phys. Rev. B **91,** 184415 (2015).
19. F. Atoneche, A. M. Kalashnikova, A. V. Kimel, A. Stupakiewicz, A. Maziewski *et al*. *Large ultrafast photoinduced magnetic anisotropy in a cobalt-substituted yttrium iron garnet,* Phys. Rev. B **81,** 214440 (2010).
20. A. V. Kimel, A. Kirilyuk, P. A. Usachev, R. V. Pisarev, A. M. Balbashov *et al*. *Ultrafast non-thermal control of magnetization by instantaneous photomagnetic pulses*, Nature **435**, 655–657 (2005).
21. F. Hansteen, A. V. Kimel, A. Kirilyuk, Th. Rasing, *Nonthermal ultrafast optical control of the magnetization in garnet films*, Phys. Rev. B. **73,** 14421 (2006).
22. C. D. Stanciu, F. Hansteen, A. V. Kimel, A. Kirilyuk, A. Tsukamoto *et al*. *All-optical magnetic recording with circularly polarized light*, Phys. Rev. Lett. **99**, 047601 (2007).
23. A. H. M. Reid, A. V. Kimel, A. Kirilyuk, J. F. Gregg, Th. Rasing, *Optical excitation of a forbidden magnetic resonance mode in a doped lutetium-iron-garnet film via the inverse Faraday effect,* Phys. Rev. Lett. **105,** 107402 (2010).
24. M. Deb, M. Vomir, J. Rehspringer, J. Bigot, *Ultrafast optical control of magnetization dynamics in polycrystalline bismuth doped iron garnet thin films*, Appl. Phys. Lett. **107,** 252404 (2015).
25. R. V. Mikhaylovskiy, E. Hendry, V. V. Kruglyak, *Ultrafast inverse Faraday effect in a paramagnetic terbium gallium garnet crystal,* Phys. Rev. B **86,** 100405 (2012).
26. T. Liu, T. Wang, A. H. Reid, M. Savoini, X. Wu *et al*. *Nanoscale Confinement of All-Optical Magnetic Switching in TbFeCo--Competition with Nanoscale Heterogeneity*, Nano Lett. **15,** 6862–6868 (2015).
27. K. Uchida, H. Adachi, D. Kikuchi, S. Ito, Z. Qiu *et al*. *Generation of spin currents by surface plasmon resonance,* Nat. Commun. **6,** 5910 (2015).
28. L. P. Pitaevskii, *Electric forces in a transparent dispersive medium.* J. Exptl. Theoret. Phys. **39**, 1450-1458 (1960).
29. J. Kim, M. Vomir, J. Bigot, *Controlling the spins angular momentum in ferromagnets with sequences of picosecond acoustic pulses*, Sci. Rep. **5**, 8511 (2015).
30. R. Gieniusz, L. Smoczyński, *Magnetostatic spin waves in (111)-oriented thin garnet films with combined cubic and uniaxial anisotropies,* J. Magn. Magn. Mater. **66,** 366–372 (1987).
31. A. I. Akhiezer, V. G. Bar'yakhtar, S. V. Peletminskii, *Spin Waves* (North-Holland, Amsterdam, 1968).
32. V. I. Belotelov, I. A. Akimov, M. Pohl, V. A. Kotov, S. Kasture *et al*. *Enhanced magneto-optical effects in magnetoplasmonic crystals,* Nature Nanotech. **6,** 370–376 (2011).